\documentclass{aastex}
\usepackage{spr-astr-addons}
\usepackage{url}

\RequirePackage{color}

\begin{document}

\title{Electromagnetic vortex beam dynamics in degenerate electron-positron astrophysical plasmas}


\author{V.I. Berezhiani\altaffilmark{1,2}, Z.N. Osmanov\altaffilmark{1,3} \& S.V. Mikeladze\altaffilmark{2}}


\altaffiltext{1}{School of Physics, Free university of Tbilisi, Tbilisi 0159, Georgia.}
\altaffiltext{2}{Andronikashvili Institute of Physics (TSU), Tbilisi \ 0177, Georgia.}
\altaffiltext{3}{E. Kharadze Georgian National Astrophysical Observatory, Abastumani 0301, Georgia.}

\begin{abstract}
For degenerate astrophysical electron-positron plasmas we have considered dynamics of electromagnetic  beams carrying angular momentum. It is found  for arbitrary level of degeneracy such a beam having the power exceeding a certain critical value breaks up into many filaments, eventually leading to the formation of stable spatial solitons keeping zero field in the center of the structure.
 
\end{abstract}



\section{Introduction}
A wide class of astrophysical objects emit the electromagnetic (EM) waves in a broad frequency range. These waves interacting with matter surrounding the objects might lead to non-trivial observational features \citep{carroll, shapiro}. It is established that under certain conditions radiation generated in astrophysical environment can acquire orbital angular momentum (twisted radiation). A series of papers has been dedicated to the study of different aspects of the mentioned problem. In particular, \cite{poam} has considered the role of twisted radiation in different aspects: influence of the Kerr black hole on angular properties of radiation; search for extraterrestrial intelligence; considering astrophysical masers as probes for inhomogeneities of the interstellar medium and the study of pulsars and quasars. It is shown that in case of the Kerr black hole the photon orbital angular momentum might be generated by the partial "absorption" of the black hole's angular momentum. This effect has been used to measure the rate of rotation of the central black hole of the active galactic nuclei (AGN) M87 \citep{tamburini}.

Recently, non-linear dynamics of EM vortex solitons in the magnetospheric electron-positron-ion relativistic plasmas of AGN has been investigated \cite{vortex}. Peculiarities of generation of vortex solitary structures and their instability was studied and possible observational signatures has been discussed.


One should emphasize that rlectron-positron (e-p) pair plasma can be generated in a wide class of compact astrophysical objects: pulsars \citep{sturrock}, AGN \citep{agn} and gamma ray bursts (GRB) \citep{grb}. Under certain conditions the gravitational collapse results in charge separation, which might lead to extremely efficient pair creation characterized by density degeneracy with the e-p number density of the order of $10^{30-37}$ cm$^{-3}$ \citep{grb,ruffini}. For this case the average distance between the closest particles becomes smaller than the de Broglie wavelength and the Fermi energy, $\epsilon_{_F} = \hbar^2(3\pi^2 n_0)^{2/3}/2m_e$ is larger than the interaction energy, $e^2n_0^{1/3}$, making the gas ideal \citep{landau}. Here we use the following notations: $\hbar$ is the Planck''s constant, $n_0$ denotes the electron/positron number density and $e$ and $m_e$ are the electron's charge and mass respectively. One can straightforwardly show that the Fermi gas is ideal if the following condition is satisfied: $n_0 >> 8m_e^3e^6/(9\pi^4\hbar^6)\simeq 6.3\times 10^{22}$ cm$^{-3}$. For even higher particle densities the Fermi energy becomes relativistic, $\epsilon_{_F} = m_ec^2\left[\left(1+R_0^2\right)^{1/2}-1\right]$, where $R_0 = (n_0/n_c)^{1/3}$ and $n_c = m_e^3c^3/(3\pi^2\hbar^3)\simeq 5.9\times 10^{29}$ cm$^{-3}$ is the critical number density. The pair annihilation time-scale for such densities is small compared to the plasma oscillation periods, and therefore collective phenomena has enough time to develop \citep{bst}.

Different aspects of EM radiation propagating in highly degenerate electron-ion as well as e-p plasma has been investigated in a series of papers \citep{bo2,gosh,bst,haas,mikab}.

In the present paper we consider non-linear dynamics of EM beams carrying orbital angular momentum propagating in highly degenerate relativistic e-p plasmas.

The paper is organised in the following way: in Sec. 2 we introduce the equations governing the self-guiding regime in the astrophysical e-p plasmas, numerically solve them and obtain results and in Sec. 3 we summarise them.

\section{Theory and discussion}

In this section we introduce the equations governing the process of self-guiding, we numerically solve them and discuss the obtained results.

By using the reductive perturbation method \citep{bs16} that the 
dynamics of EM beams propagating in
degenerate e-p plasma is governed by the nonlinear Schrodinger equation
(NSE) with saturating nonlinearity which in dimensionless form reads

\begin{equation}
2i\frac{\partial A}{\partial z}+\nabla _{\perp }^{2}A+f\left( \left\vert
A\right\vert ^{2}\right) A=0  \label{1}
\end{equation}%
with

\begin{equation}
f\left( \left\vert A\right\vert ^{2}\right) =1-\frac{\left( 1-\left\vert
A\right\vert ^{2}\right) ^{3/2}}{\left( 1-\left\vert A\right\vert
^{2}d\right) ^{1/2}}  \label{2}
\end{equation}

Here $A$ is the slowly varying amplitude of the circularly polarized vector
potential 

\begin{equation}
\frac{e\bf{A}}{m_ec^2R_0} \mathbf{=}(1/2)\left( 
\widehat{\mathbf{x}}+i\widehat{\mathbf{y}}\right) A\exp (-i\omega
_{0}t-k_{0}z)+c.c., \label{eA}
\end{equation}
where $\widehat{\mathbf{x}}$ and $\widehat{\mathbf{y}}$
are unite vectors directed across the propagation of the EM beam $z$ . The
field frequency $\omega _{0}$ and wave vector $k_{0}$ satisfy the dispersion
relation $\omega _{0}^{2}=k_{0}^{2}c^{2}+2\omega _{e}^{2}/\Gamma _{0}$ , $%
\omega _{e}=\left( 4\pi e^{2}n_{0}/m_{e}\right) ^{1/2}$ - is the plasma
frequency and $\Gamma
_{0}=\left( 1+R_{0}^{2}\right) ^{1/2}$ is the generalized relativistic
factor where $R_{0}=\left( n_{0}/n_{c}\right) ^{1/3}$. \ The dimensionless
coordinates reads as $z=\left( 2\omega _{e}^{2}/c\omega _{0}\Gamma
_{0}\right) z$, $\mathbf{r}_{\perp }=\left( \sqrt{2}\omega _{e}/c\sqrt{%
\Gamma _{0}}\right) \mathbf{r}_{\perp }$ and the operator $\nabla _{\perp
}^{2}$ is the Laplacian in the $\mathbf{r}_{\perp }=(x,y)$ plane. In Eq. (2) 
$d=R_{0}^{2}/\left( 1+R_{0}^{2}\right) <1$ measures a level of degeneracy.
For the weakly degenerate case $\left( R_{0}<<1\right) $ $d\simeq R_{0}^{2}$
while for the relativistic degeneracy $\left( R_{0}>>1\right) $ $d\rightarrow 1$. 
In the system of Eqs. (1)-(2) it is assumed that plasma is highly
transparent $\omega _{e}/\omega _{0}<<1$ and $\lambda <<L_{\perp
}<<L_{\parallel }$ where $\lambda \approx 2\pi c/\omega _{0}$ is the
wavelength of the EM radiation, $L_{\perp }$ and $L_{\parallel }$ are the
characteristic longitudinal and transverse spatial dimensions of the EM
beam. This system describes the dynamics of strong amplitude narrow EM beams
in e-p plasma with the arbitrary strength of degeneracy.

In an unmagnetized e-p plasma the EM pressure is equal for both the electrons
and positrons and consequently modification of plasma density takes place
without producing the charge separation. For the normalized
electron/positron density $N=N/n_{0}$ we have the following expression

\begin{equation}
N=\frac{\left( 1-\left\vert A\right\vert ^{2}\right) ^{3/2}}{\left(
1-\left\vert A\right\vert ^{2}d\right) ^{1/2}}  \label{3}
\end{equation}%
From Eq.(3) it follows that our considerations remain valid provided that $%
\left\vert A\right\vert <1$; the plasma density decreases in the area of the EM
field localization and if at a certain point of this area one has $\left\vert
A\right\vert \rightarrow 1$, then the plasma density becomes zero ($%
N\rightarrow 0$). Therefore, at that point the cavitation takes place \cite{bo2}. Here
we would like to remark that the condition $\left\vert A\right\vert <1$ does
not necessarily imply that the strength of the EM field is relativistic. In dimensions
this condition reads as $e\left\vert A\right\vert \mathbf{/}\left(
m_{e}c^{2}\right) <R_{0}$ and for a weakly degenerate case ($R_{0}<<1$) the
cavitation can take place even if the strength of the EM field is weakly relativistic. At
this end we would like to emphasize that values of the parameter $R_{0}$ are
bounded from below and can not be taken to be zero. Indeed, for the
degenerate plasma the average energy of e-p particle interaction should be
less the Fermi energy. This condition implies that plasma density should be $%
n_{0}\geq e^{6}m_{e}^{3}/\hbar ^{6}=\allowbreak 6.\,7\times 10^{24}cm^{-3}$
\ and consequently $R_{0}>>0.02$ ($d>>4\times 10^{-4}$).

\begin{figure}
  \centering {\includegraphics[width=8.4cm]{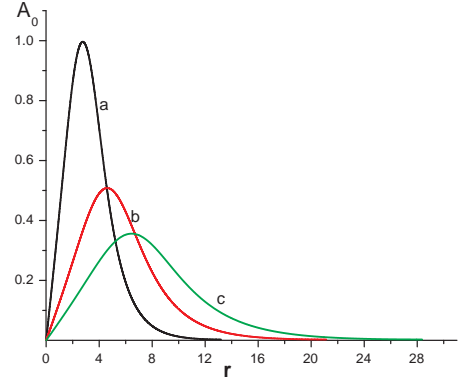}}
  \caption{Here we show the plots of ground state solutions versus the propagation constant for the strongly degenerate state $d = 0.5$ with $m = 1$.}\label{fig1}
\end{figure}

Based on the system of Eq.(1) in \citep{bs16} the authors demonstrated that under well
defined conditions e-p plasma supports existence of fundamental (noddles)
stable 2D solitonic structures for arbitrary values of the degeneracy parameter $%
d$. The EM beam with above certain critical power can be trapped in
self-guiding regime of propagation and subsequently formation solitonic
structures takes place.

At first we consider 2D solitary wave solutions carrying vortices. On
assuming that solutions in polar coordinates are of the form $A=A_{0}\left(
r\right) \exp \left( im\theta +ikz\right) $, where $r=\sqrt{x^{2}+y^{2}}$ , $%
\theta $ is the polar angle, Eq.(1)\ reduces to an ordinary differential
equation the real valued amplitude $A_{0}$

\begin{equation}
\frac{d^{2}A_{0}}{dr^{2}}+\frac{1}{r}\frac{dA_{0}}{dr}-kA_{0}-\frac{m^{2}}{%
r^{2}}A_{0}+A_{0}\left( 1-\frac{\left( 1-A_{0}^{2}\right) ^{3/2}}{\left(
1-A_{0}^{2}d\right) ^{1/2}}\right) =0  \label{4}
\end{equation}%
where $A_{0}$ is the real valued amplitude, $k$ is nonlinear propagation
constant, and $m\left( \neq 0\right) $\ is an integer known as the
topological charge of the vortex.

For $0<k<1$ Eq.(4) admits an infinity of localized discrete bound states $%
A_{0n}\left( r\right) $ with $A_{0n}\left( 0\right) =0$ and $A_{n}\left(
r\rightarrow \infty \right) \rightarrow 0$ with with the following
asymptotic behavior: $A_{0n}\left( r\rightarrow 0\right) \sim r^{\left\vert
m\right\vert }$ and $A_{0n}\left( r\rightarrow \infty \right) \sim \exp
\left( -r\sqrt{k}\right) /\sqrt{r}$. Here $n$ denotes number of zeros of
eigenfunction for $r\neq 0$. In what follows we consider the lowest order,
the ground state solutions, which have the node at the origin $r=0$, reaches
a maximum, and then monotonically decrease with increasing $r$. Such
solutions are obtained numerically applying the shooting code for different
level of degeneracy parameter $d$.

\begin{figure}
  \centering {\includegraphics[width=8.4cm]{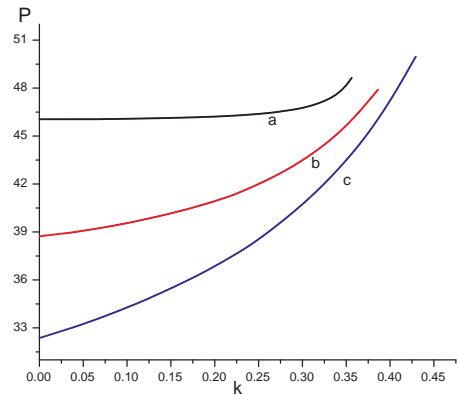}}
  \caption{For $m = 1$ the dependence of power on the propagation constant is presented for different values of degeneracy, $d = (0.01; 0.05; 0.9).$}\label{fig2}
\end{figure}

\begin{figure}
  \centering {\includegraphics[width=8.4cm]{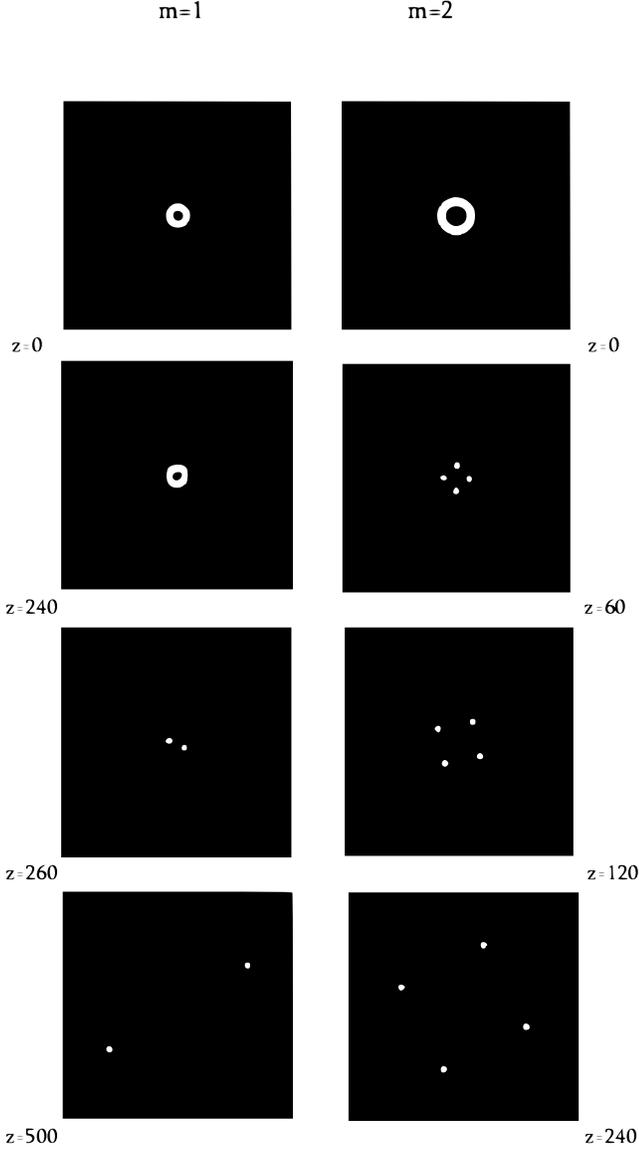}}
  \caption{Here we show the contours of constant intensity beams for $m = 1,2$ and $d = 0.5$.}\label{fig3}
\end{figure}

For $m=1$\ in Fig. 1 we present the profiles of ground state solutions
versus the propagation constant, $k$, for the strongly degenerate case $d=0.5$ .
Similar profiles of the solitons can be contained for arbitrary values of $d$
$\left( <1\right) $. The maxima of the field amplitudes $A_{m}$ $<<1$ for $%
k\rightarrow 0$ while for $k\rightarrow k_{cr}(d)<1$, $A_{m}\rightarrow 1$,
i.e. the plasma cavitation takes place. Here $k_{cr}(d)$ depends on the
level of degeneracy $d,$ \ and, for instance $k_{cr}(0.01)\simeq 0.44$, $%
k_{cr}(0.5)\simeq 0.37$\textbf{. }Similar behavior of the solutions can be
obtained for vortices with higher charge ($m=2,3,...$); corresponding
figures are not displayed here.

The power of single charged ($m=1$) EM\ beam trapped in the vortex soliton
modes $\left( P=2\pi \int_{0}^{\infty }rA_{0}^{2}dr\right) $ is a growing
function of $k$. Dependence of $P$\ on the propagation constant $k$\ for
different level of plasma degeneracy $d=0.01;0.5;0.9$ is exhibited in Fig.2.
For $k\rightarrow 0$ ($A_{m}\rightarrow 0$), $P\rightarrow P_{cr}\left(
d\right) $ where $P_{cr}\left( d\right) $ is a critical power. In case of
weak degeneracy $\left( d=0.01\right) $ the critical power $P_{cr}(0)\simeq
32.1$ while for relativistic degeneracy cases $P_{cr}(0.5)\simeq 38.6,$ $%
P_{cr}(0.9)\simeq 46.1$. Thus, the necessary condition to form the vortex
soliton is that the EM beam power must exceed $P_{cr}(d)$ while an upper
bound of the power is related to the plasma cavitation that is taking place
at $k\rightarrow k_{cr}$ ($A_{m}\simeq 1$).

In dimensional unites the critical power is expressed as 
\begin{equation}
P_{d}\simeq 0.17\left( \omega /\omega _{e}\right) ^{2}\Gamma
_{0}R_{0}^{2}P_{cr}\left( d\right) \left[ GW\right]  \label{5}
\end{equation}%
where $P_{d}$ is the critical power measured in $GW$. For $d=0.01$ ($%
R_{0}\simeq 0.1)$ and, for instance $d=0.5$ ($R_{0}=1$) the corresponding
densities of plasma are respectively $n_{0}\simeq 5.96\times 10^{26}cm^{-3}$
and $n_{0}\simeq 5.96\times 10^{29}cm^{-3}$. Since plasma is assumed to be
transparent photon energies of EM beam should be in hard $X$- ray band $%
\left( \hbar \omega >>(1KeV-29KeV)\right) $ while for the critical powers we
have $P_{d}=(5.4-6.6)GW$.

The stability of the vortex soliton solutions of NSE have been studied in
the past for different king of saturating nonlinearity
\citep{firth,skarka}. It is well established that though $dP/dk>0$ and the Vakhiton and
Kolokolov criteria guaranties stability of solutions against small \ radial
perturbations , for the symmetry breaking azimuthal perturbations the
solitons are unstable. The instability causes the to breakup of the
structure into multiple filaments while number of filaments are usually $2m$
. The filaments must conserve total angular momentum $\left( \sim mP\right) $
and also they can not fuse to due to topological reasons. These filaments
carry zero topological charges ($m=0$), they can eventually spiral about
each other or fly off tangentially to the initial ring generating stable
solitonic 2D structures. Our simulations demonstrate that similar scenario
of instability development takes place for the degenerate e-p plasma
described by Eqs.(1)-(2). \ 

We carried out the numerical simulations to investigate the dynamics of EM
vortex beam with parameters and shapes far from ground state vortex
solitons. For this purpose an input vortex beam is assumed to be Gaussian $%
A(z-0,x,y)=(x+iy)^{m}A_{1}\exp (-r^{2}/2D^{2})$. The radial profile of beam
intensity being zero at $r=0$, reaches a maximum $A_{\max
}=A_{1}D^{m}m^{m/2}/\exp $ $\left( m/2\right) $ at $r_{\max }=D\sqrt{m}$ and
then exponentially decays with increasing $r$. Here $D$ is the
characteristic width of the structure and $A_{1}$ is a constant which (along
with $D$) determines the EM beam power $P=\pi A_{1}^{2}D^{2\left( m+1\right)
}m!$. For numerical simulations it is comfortable to characterize input beam
by its power and amplitude $A_{\max }<1$. If the power of the beam less than
the critical one $P<P_{cd}$ \ the beam diffracts while for $P>P_{cd}$ \
breakup of the structure into filaments takes place for all levels of
degeneracy parameter $\left( 0<d<1\right) $. In Fig. 3 contours of constant
intensity of the beams are displayed for $m=1,2$ and $d=0.5$. For both cases
we assume that $P=42$ and $A_{\max }=0.3$ implying that the vacuum
diffraction length $(z_{dif}=D^{2})$ for single and double charges beams are
respectively $z_{dif}\simeq 55$ and $z_{dif}\simeq 40$. One can see that in
few diffraction length the single charged vortex beam breaks up into two
while double charged beam into four filaments. These filaments carry zero
topological charge ($m=0$), flying off tangentially to the initial ring like
distribution of EM field intensity with subsequent formation of stable 2D
spatial solitons.

\section{Conclusion}

Critical power of the EM beam carrying angular momentum has been found. It is shown that the vortex soliton solution with power above the critical value turns out to be unstable against symmetry breaking perturbations, leading to the formation of stable soliton solutions with zero topological charge, keeping zero field in the centre of the structure.

We demonstrated that in general, the EM beam with an initial Gaussian shape, having the power less than the critical one diffracts monotonically. However, if the power is overcritical the vortex beam breaks into filaments with the subsequent formation of stable solitary structures with an arbitrary level of degeneracy.

Self-guiding structure which is composed of beamlets are running away tangentially from the initial ring of field distribution, leaving the zero-field at the centre. In the course of propagation these beamlets form stable 2D spatial solitons.

\bibliographystyle{spr-mp-nameyear-cnd}

\end{document}